\documentstyle[prl,aps,twocolumn]{revtex}
\begin{document}
\draft
\title{ Dynamics of a Vortex in Two-Dimensional Superfluid He3-A: 
Force Caused by the $l$-Texture}
\author{Hiroshi Kuratsuji${}^1$ and Hiroyuki Yabu${}^2$}
\address{ ${}^1$Department of Physics,
               Ritsumeikan University-BKC,
               Kusatsu City, 525-77, Japan \\
${}^2$Department of Physics,
               Tokyo Metropolitan University,
               Hachioji, Tokyo 192, Japan }
\date{\today}
\maketitle
\begin{abstract}
Based on the Landau-Ginzburg Lagrangian, the dynamics of a vortex is 
studied for superfluid He3-A characterized by the ${\bf l}$-texture. 
The resultant equation of motion for a vortex leads to the Magnus-type 
force caused by the ${\bf l}$-texture. The force is explicitly written 
in terms of the mapping degree from the compactified 2-dimensional plane 
to the space of ${\bf l}$-vector, which reflects the quantitative 
differences of vortex configurations, especially the Mermin-Ho and 
Anderson-Toulouse vortices. The formulation is applied to anisotropic 
superconductors in which the Hall current is shown to incorporate changes 
between vortex configurations. 
\end{abstract}
\pacs{PACS number: 67.40.Vs}

{\it Introduction}:
The study of superfluid He3 is still in the limelight in condensed matter 
physics. Among a variety of problems, great efforts have been devoted 
to the investigation of vortices in superfluid He3-A phase
\cite{Volovik1,Volovik87}.  
The order parameter for He3-A is anisotropic and characterized by 
``${\bf l}$-texture'' that resemble the ``director'' in the liquid 
crystal. This feature gives rise to a variety of phenomena 
in vortex physics\cite{Volovik87,Sonin87}. 

The purpose of this letter is to give a new view of vortex dynamics, 
specifically, for the non-singular ${\bf l}$-texture vortex inherent 
in the ${\bf l}$-texture. In order to carry out this, we derive the 
equation of motion for a single vortex from the time-dependent 
Landau-Ginzburg-type Lagrangian which is a generalization of that 
for ordinary superfluid \cite{Feynman,Kuratsuji92}. 
Here it is notable that, if one uses the vector-like order parameter
including the ${\bf l}$-texture, the Lagrangian is reduced to the one 
for the spin field and become similar with that for the ferromagnet model 
used in the study of the magnetic vortex motion \cite{Ono94,Kuratsuji96}: 
an alternative form has been proposed previously in \cite{Cross77}, 
but the one adopted in the present paper has much simpler form, 
from which the equation of motion can be easily derived. 

The resultant equation of motion for a single vortex shows a specific 
non-dissipative Magnus-type force familiar in the ordinary superfluid 
or superconductivity \cite{Nozieres66,Bardeen65}. 
A remarkable consequence is that the strength of this force is explicitly 
given by the mapping degree (or topological invariant) from two-dimensional 
space to the space of {\bf l}-vector, which reflects the boundary condition 
of the vortex at infinity (or at the wall of containers). 
Especially, two typical vortices can be incorporated, the Mermin-Ho (MH) 
and Anderson-Toulouse (AT) vortices \cite{Anderson,Ho76}. 

As an application to anisotropic superconductors, we consider the Hall 
currents, in which this peculiar feature is shown to be reflected naturally. 

\bigskip 

{\it Landau-Ginzburg Lagrangian for He3-A}: 
In the present paper, our argument is restricted to two-dimensional superfluid 
uniform in the z-direction and the zero temperature case. Let $\psi$ be the 
vector-type order parameter describing the Cooper pair in He3-A, which has 
the form\cite{foot0}
\begin{equation}
   \psi = \Delta_0 e^{i\gamma}({\bf e_1}- i{\bf e_2})
\label{two}
\end{equation}
where $ {\bf e_1}, {\bf e_2} $ are the orthogonal unit vector fields 
that make the triad with $ {\bf l} = {\bf e_1}\times {\bf e_2} $, 
and $\gamma$ is the rotation angle around the ${\bf l}$-vector\cite{Ohmi95}. 
The ${\bf l}$-vector represents the orbital angular momentum of 
the Cooper pair.
These triad vectors are represented in terms of the angular 
variables\cite{Ohmi95}; 
\begin{eqnarray}
{\bf l} & = & (\sin\theta\cos\phi, \sin\theta\sin\phi, 
\cos\theta) \nonumber \\
{\bf e_1} & = & (\cos\theta\cos\phi, \cos\theta\sin\phi, 
-\sin\theta), \nonumber \\
{\bf e_2} &  = & ( -\sin\phi, \cos\phi, 0) 
\label{three}
\end{eqnarray}
where $(\theta, \phi)$ represent the polar angles of the ${\bf l}$-vector. 
The Lagrangian for the vector order parameter is given as an extension 
from that for the usual superfluid\cite{Feynman}: 
\begin{eqnarray}
   L &  = & \int \{{i\hbar \over 2}(\psi^{\dagger}{\partial \psi 
         \over \partial t} - c.c)- H(\psi, \psi^{\dagger})\}d^2x 
                \nonumber \\
     &  \equiv & L_C - L_H 
\label{one}
\end{eqnarray}
where the first term is regarded as a variant of the geometric 
phase\cite{Shapere} and the second term is nothing but the Hamiltonian 
$L_H$. By the requirement of the rotational invariance, the general form 
of $L_H$ is given by
\begin{eqnarray}
L_H & = & \int{d^2x} [K_1 \nabla_i \psi^{\dagger}_j
\nabla_i \psi_j + K_2 \nabla_i \psi^{\dagger}_i  
\nabla_j \psi_j  \nonumber \\  
    & +  &  K_3\nabla_i\psi^{\dagger}_j\nabla_j\psi_i] 
\label{eQone}
\end{eqnarray}
For the coefficients $K_{1,2,3}$, we adopt the results in the critical 
region ($T \sim T_c$): $K_1=K_2=K_3=\frac{1}{5} N(0) \xi_0^2$ 
($\xi_0=\sqrt{\frac{7\xi(3)}{48\pi^2}}$) in the weak-coupling 
approximation\cite{Volovik87,Brinkman78}. 

Substituting (\ref{three}) into (\ref{one}), we get the polar 
representation of the canonical term $L_C$: 
$ i\hbar \psi^\dagger {\dot\psi} - c.c. 
  = 2\hbar\Delta_0^2 ({\dot\gamma} -\cos\theta{\dot\phi}) $. 
Using the gauge fixing condition $\gamma =\phi$, 
it becomes 
\begin{equation} 
 L_C = -2 \hbar \Delta_0^2 \int (1 - \cos\theta) {\dot\phi} d^2x 
\label{five}
\end{equation}
In the previous papers\cite{Ono94,Kuratsuji96}, this form was used 
in the continuous Heisenberg model. The angular form of the Hamiltonian 
term will be discussed below.
\bigskip 

{\it Equation of motion for a coreless vortex}:
 Now let us consider a single vortex located at the origin, 
for which the angular variable $\phi$ is taken to be $\phi =\arctan(y/x)$, 
whereas the profile function $\theta$ obeys the boundary conditions 
$\theta(0)=0$ and, 
1) $\theta(\infty) =0$ (MH vortex) or 
2) $\theta(\infty) ={\pi \over 2}$ (AT vortex). 
They have a finite size near the origin (``core''), 
and the ${\bf l}$-vector directs upward inside the core, 
while it is ``planer'' (MH vortex) or directs downward 
(AT vortex) at infinity (or at the container wall). 

We introduce the collective coordinate ${\bf X}(t) = (X, Y)$ 
to describe the dynamics of a vortex and replace the space arguments 
of the angular variables from $ {\bf x} $ to ${\bf x} - {\bf X}(t) $. 
With the chain rule, ${\partial \phi \over \partial t} 
= {\partial \Phi \over \partial {\bf X}}\dot{{\bf X}}$, 
the canonical term $L_C$ becomes 
\begin{equation}
 L_C = -2\hbar\Delta_0^2\int(1 - \cos\theta)
\nabla\phi \cdot\dot{\bf X}d^2x  
\label{six}
\end{equation}
from which one sees that the momentum density ${\bf p}$ is defined to be 
canonically conjugate with $\dot{\bf X}$ such that 
\begin{equation}
     {\bf p} =\rho' {\bf v} \equiv M \Delta_0^2{\bf v}
\label{seven}
\end{equation}
where $M$ is a parameter with mass dimension and the velocity field 
${\bf v}$ is defined by 
\begin{equation}
{\bf v} = -{\hbar \over M} (1 - \cos\theta)\nabla \phi. 
\label{eight} 
\end{equation}
This velocity field does not bear any singularities near the origin, 
and such vortices are called ``soft-core''. 

The explicit form of the Hamiltonian term is given in \cite{Ohmi95}, and 
that can be classified as $L_H =T_2+T_1+T_0$, where $T_n \propto {\bf v}^n$. 
The explicit form of $T_2$ is given by 
\begin{eqnarray}
T_2 & = & [{1\over 2}\rho^\|({\bf l}{\bf v})^2 
+ {1\over 2}\rho^{\bot}({\bf l}\times {\bf v})^2] \nonumber \\
 & = & {1\over 2}[\rho^{\bot}{\bf v}^2 
      + (\rho^\| - \rho^{\bot})({\bf l}{\bf v})^2]
\end{eqnarray}
where, in \cite{Ohmi95}, the mass densities are 
$\rho^\bot =\rho^\| =4(m/\hbar)^2 K_1 \Delta_0^2$. 
($m$ is twice the He3 mass.) In two-dimensional case, 
direct calculation shows ${\bf l}{\bf v} = 0$, and $T_2$ becomes 
\begin{equation}
T_2 ={1\over 2} \int \rho {\bf v}^2 dx^2 
\end{equation}
where $\rho =\rho^\bot$. Note that the density $\rho$ should coincide 
with $\rho'$ in (\ref{seven}), so that the parameter $M$ should be fixed 
to be $4(m/\hbar)^2 K_1 =M$. It should be noted that the $T_2$ has just 
the same form with the kinetic energy for the superfluid. 

As for the remaining term, the explicit form of $T_1$ is 
\begin{equation}
T_1 =\int {\hbar \rho^\| \over 2m} 
     [ {\bf v}(\nabla \times {\bf l}) 
      -2 ({\bf v}{\bf l}) {\bf l}(\nabla \times {\bf l}) ] d^2x
\end{equation}
and the second term of it also vanishes in two-dimensional case 
(${\bf l}{\bf v}=0$) and the first term will not contribute to the 
Magnus-type force because it becomes the total divergence after ${\bf v}$ 
is replaced by the constant velocity ${\bf U}$. The ${\bf v}$-independent 
term $T_0$ does not contribute to the Magnus-type force, and we can drop it. 
(For the explicit form of $T_0$, see \cite{Ohmi95,foot2})

Let us consider the equation for the vortex center ${\bf X}$ 
in the presence of the background uniform superflow ${\bf U}$.
In that case, the kinetic term $T_2$ becomes  
\begin{eqnarray}
 T_2 & = & {1\over 2}\int\rho({\bf v} + {\bf U})^2 d^2x 
\nonumber \\
  & = &  {1\over 2}\int\rho {\bf v}^2d^2x 
        + \int\rho{\bf v} \cdot{\bf U}d^2x 
        + {1\over 2}\int\rho{\bf U}^2 d^2x  
\label{nint}
\end{eqnarray}
The equation of motion for a vortex is given by the Euler-Lagrange equation 
\begin{equation}
{d \over dt}{\partial L \over \partial \dot{\bf X}}
            - {\partial L \over \partial {\bf X}} = {\bf F}_{ex}
\label{eighttwo}
\end{equation}
The resultant equation is rewritten ${\bf F}_C + {\bf F}_T = {\bf F}_{ex}$ 
where $ {\bf F}_C $ and $ {\bf F}_T $ are given by 
$ {\bf F}_C \equiv 
{d \over dt}{\partial L_C \over \partial \dot{\bf X}} 
            - {\partial L_C \over \partial {\bf X}} $ 
and 
$ {\bf F}_T \equiv -{\partial T_2 \over \partial {\bf X}} $. 
The external force ${\bf F}_{ex}$ represents the force for balance that may 
come from the other types of non-dissipative force as well as dissipative 
force. 
The two-dimensional space integral can be performed in the moving frame; 
${\bf x}' = {\bf x} - {\bf R}(t)$ and ${\bf X}' = {\bf X} - {\bf R}(t)$, 
where ${\bf R}(t)= {\bf U}t$ represents the center of mass of the whole 
system\cite{foot3}. The canonical term (\ref{six}) is written by 
\begin{equation}
L_C = 2m\Delta_0^2\int {\bf v}({\bf x}' - {\bf X}'(t))\dot{\bf X}d^2x'.
\label{twone}
\end{equation}
Using eq. (\ref{eighttwo}), we get the force from (\ref{twone}): 
\begin{eqnarray}
{\bf F}_C & = & 2M\Delta_0^2{\bf k} \times (\dot{\bf X} - {\bf U})
[\int (\nabla \times {\bf v})_zd^2x']  \nonumber \\
        & & -{1\over 2}(2M)\Delta_0^2
({\bf k}\times {\bf U})\int(\nabla\times {\bf v})_zd^2x'
\label{twthree}
\end{eqnarray}
where the vector $ {\bf k} $ is the z-directed unit vector. 

Let us turn to  ${\bf F}_T$. The direct calculation shows that only the 
second term $T'$ in (\ref{nint}) contributes for it. 
In the moving frame, $T'$ becomes 
\begin{equation}
T' = \int \rho{\bf v}({\bf x}' - {\bf X}'(t))\cdot{\bf U}d^2x' 
\label{twtwo}
\end{equation}
which results in \cite{foot4}
\begin{equation}
{\bf F}_T = -{\partial T \over \partial {\bf X}}=
[{1\over 2}(2M)\Delta_0^2 \int (\nabla \times {\bf v})_zd^2x']
({\bf k}\times {\bf U})
\label{twtfive}
\end{equation}
By summing up (\ref{twthree}) and (\ref{twtfive}), we get the balance of 
forces: 
\begin{equation}
   2M\Delta_0^2[\int({\partial v_y \over \partial x} 
   - {\partial v_x \over \partial y})d^2x]
{\bf k}\times (\dot{\bf X} - {\bf U}) = {\bf F}_{ex}.  
\label{twseven}
\end{equation}
We note that the Galileian invariance is guaranteed in this formula. 
Eq.(\ref{twseven}) is the Magnus-type force in superfluid He3-A. 
When ${\bf U} = 0$, eq.(\ref{twseven}) becomes 
\begin{equation}
{\bf F}_C = [2m\Delta_0^2\int(\nabla \times {\bf v})_zd^2x]
({\bf k} \times \dot{\bf X})
\end{equation}
that is the same as for a vortex in ferromagnet\cite{Kuratsuji96}. 

To examine the significance of the strength of the Magnus-type force, 
we consider the integral in eq.(\ref{twseven}) 
\begin{equation}
  \sigma = \int_{R^2}({\partial v_y \over \partial x} 
         - {\partial v_x \over \partial y})d^2x 
\label{eleven}
\end{equation} 
It can be easily proved that the integrand of $\sigma$ is the continuous 
vorticity: 
\begin{equation}
\omega \equiv \nabla \times {\bf v} 
       ={\hbar/M} \sin\theta \nabla \theta \times \nabla \phi 
\label{twelv}
\end{equation}
which can be rewritten in terms of the ${\bf l}$-vector as
\begin{equation}
     \omega ={\hbar \over M}
             {\bf l}\cdot( {\partial {\bf l} \over \partial x}
                 \times {\partial {\bf l} \over \partial y}) 
\label{thirt}
\end{equation}
This relation is the nothing but Mermin-Ho relation \cite{Ho76}. Thus the 
$\sigma$ can be described as the integral over a part of the ${\bf l}$-space 
or two dimensional sphere 
\begin{equation}  
\sigma = \int_S \sin\theta d\theta\wedge d\phi 
\label{fourt}
\end{equation}
On account of the boundary condition for the ${\bf l}$-vector, $\sigma$ 
serves as the topological invariant\cite{Volovik87,Mermin79}; 
a) MH vortex, $ l_3(\infty) = 0 $, 
 ($ \theta = {\pi \over 2} $ ) and 
b) AT vortex, $l_3(\infty) = -1 $, ($ \theta = \pi $), 
for which the images S become the ``hemisphere'' and the ``whole sphere'' 
in the ${\bf l}$-space each other. Correspondingly, eq.(\ref{fourt}) yields 
the mapping degree: ${\bf R}^2 \rightarrow {\rm S}^2/2$ for the case a) 
and ${\bf R}^2 \rightarrow {\rm S}^2$ for the case b). Thus we have 
$\sigma^a = n/2\times 4\pi$ and $\sigma^b = n \times 4\pi$ respectively, 
where $n$ denotes the integer. Finally, the force is written by \cite{foot5}
\begin{equation}
{\bf F}_C  = 2M\Delta_0^2\sigma^i {\bf k} \times (\dot{\bf X}- {\bf U})
\label{fift}
\end{equation}
where $i = a,b$ for MH and AT vortices each other. Eq.(\ref{fift}) is 
the first main consequence of the present paper. 

The formula (\ref{fift}) implies an interesting feature; if there may occur 
a discontinuous change in topology of vortex triggered by some rearrangement 
of the boundary condition, the magnitude of the force changes according to 
this formula. In the case of topological change between the MH and AT vortices 
mentioned above, the ratio of the magnitude is given by 
\begin{equation}
{\vert{\bf F}^a \vert \over \vert {\bf F}^b \vert } 
= {\sigma^a \over \sigma^b}
= {1\over 2} 
\label{sixt}
\end{equation}
Such a possibility of topology change is a characteristic of the soft-core 
vortex inherent in the ${\bf l}$-texture of He3-A and this feature cannot be 
expected for the vortex in He3-B phase and the s-wave superconductors. 

\bigskip

{\it An application to anisotropic superconductor}: 
We shall now address a possible effect of the Magnus-type force caused by 
the ${\bf l}$-texture in superconductors described by a vector-type order 
parameter; The heavy electron superconductors may be considered as a possible 
candidate for such an anisotropic superconductor. Here we discuss about the 
specific feature of the Hall current derived from the characteristic 
properties of the ${\bf l}$-texture. The electromagnetic field ${\bf A}(x)$ 
is introduced in gauge-invariant way by the replacement of the kinetic term: 
$\nabla \psi \rightarrow (\nabla-{ie^{*}\over \hbar c}{\bf A})\psi$. 
($ e^{*} \equiv 2e $ denotes the charge of the Cooper pair.) 
Now we consider the case where the electromagnetic field satisfies 
the condition: the constant magnetic flux 
$\Phi \equiv {e^{*} \over \hbar c}\int{\bf A}d{\bf x} $ 
is penetrating the vortex core. To realize it, it is enough to take
\begin{equation}
{e^{*} \over \hbar c}{\bf A} = rf(r)\nabla \phi 
\label{sevt}
\end{equation}
where $\phi =\tan^{-1}(x/y)$. 

Eq.(\ref{twseven}) can be easily extended for the case that the 
electromagnetic field exists, and the balance of force is modified to be 
\begin{equation}
2m\Delta_0^2 \sigma {\bf k} \times( \dot{\bf X}- {\bf U}) 
 - 2m\Delta^2\eta {\bf k} \times {\bf U} = {\bf F}_{ex} 
\label{twsix}
\end{equation}
where $ \eta = {\Phi \over 2} $. The second term in eq.(\ref{twsix}) 
originates in the presence of the magnetic flux and it violates the Galileian 
invariance. Using the relation (\ref{twsix}), the Hall current can be derived 
in the ideal limit of no dissipative forces\cite{Dorsey92} (from the electric 
current associated with the Cooper pair ${\bf J_s} = e^{*}\Delta_0^2{\bf U}$ 
and the Faraday's law; ${\bf E} = - \dot{\bf X}\times {\bf B} $). 
After simple calculations, we get the Hall current: 
\begin{equation}
{\bf J}_s = \sigma_{xy}({\bf E}\times {\bf k}) 
\label{tweight}
\end{equation}
where the Hall conductance $\sigma_{xy}$ turns out to be 
\begin{equation}
\sigma_{xy} = {\sigma \over \sigma + \eta}
{e^{*} \Delta_0^2 \over B}  
\label{twnine}
\end{equation}
The relation (\ref{tweight}) is the second main result of the present paper.
Eq.(\ref{twnine}) shows that the Hall conductance is determined from the ratio 
${\sigma / \eta}$, where $\sigma$ depends on the boundary condition at the 
infinity. It should be noted that the Hall conductance incorporates the 
difference of topology because $\sigma$ depends on the boundary condition at 
the infinity.  Hence, if the topological change occurs in superconductors, 
it should reflect on the Hall current. In particular, for the transition 
between the MH and AT vortices, the ratio of the current is given by 
\begin{equation}
{\sigma_{xy}^a \over \sigma_{xy}^b} = {\sigma^a \over \sigma^b}
{2\sigma_b + \Phi \over 2\sigma_a + \Phi} 
\label{thirty}
\end{equation}
It may provide us an observational method to detect the difference between 
two vortex states. In real anisotropic superconductors, some modifications 
may be caused by the possible magnetic moment that the Cooper pair carries 
due to the p-wave pairing, but more details may be needed to treat this 
problem at hand\cite{Volovik85}.


\begin{thebibliography}{10}
\bibitem{foot2}
This term is written in the form of the Heisenberg spin Hamiltonian, 
$  H_S  = \sum_{i=1}^3 (\nabla l_i)^2 $. See e.g. \cite{Ono94}.
\end{thebibliography}
\end{document}